\documentclass[12pt]{iopart}

\usepackage{iopams}  
\usepackage{graphicx}

\newcommand{\disorder}[1]{\mu_{\tilde{{#1}}}}
\newcommand{\wrap}[1]{\left({#1}\right)}
\newcommand{\ii}{\rmi} 
\newcommand{\ee}{\rme}
\newcommand{\edge}[2]{{\langle {#1}, {#2}\rangle}}
\newcommand{\coveredge}[2]{\edge{#1}{\tilde{#2}}}
\newcommand{\RC}[2]{\psi_{\coveredge{#1}{#2}}}
\newcommand{\ZN}{\mathrm{Z}_N}

\begin{document}

\title{Integrability as a consequence of discrete holomorphicity: 
the $\ZN$ model}

\author{I T Alam$^1$ and M T Batchelor$^{1,2}$}

\address{$^1$ Department of Theoretical Physics, 
Research School of Physics and Engineering, 
The Australian National University, Canberra ACT 0200, Australia}
\address{$^2$ Mathematical Sciences Institute, 
The Australian National University, Canberra ACT 0200, Australia}

\ead{Imam.Alam@anu.edu.au}
\ead{Murray.Batchelor@anu.edu.au}


\begin{abstract}
It has recently been established that imposing the condition of 
discrete holomorphicity on a lattice parafermionic observable 
leads to the critical Boltzmann weights in a number of lattice models.
Remarkably, the solutions of these linear equations also solve the 
Yang-Baxter equations.  We extend this analysis for the $\ZN$ 
model by explicitly considering the condition of discrete 
holomorphicity on two and three adjacent rhombi. For two rhombi this
leads to a quadratic equation in the Boltzmann weights and for three
rhombi a cubic equation. The two-rhombus equation implies the inversion
relations. The star-triangle relation follows from the three-rhombus
equation. We also show that these weights are self-dual as a 
consequence of discrete holomorphicity.
\end{abstract}

\section{Introduction}

In some remarkable developments, a surprising connection has been
uncovered between the notions of discrete holomorphicity and 
Yang-Baxter integrability \cite{RC,IC,C,LR,IR}. On the one hand, 
a parafermionic observable on the lattice is discretely holomorphic, 
i.e., the observable satisfies a version of the discrete Cauchy-Riemann
equations. This requirement leads to a set of linear equations 
which can be solved to yield the Boltzmann weights of the model 
at criticality. The surprise is that, on the other hand, these are 
the same Boltzmann weights that follow by solving the star-triangle or
Yang-Baxter equations.
This connection has been observed by Rajabpour and Cardy \cite{RC} for
the $\ZN$ model and by Ikhlef and Cardy \cite{IC} for the Potts model,
the dilute $O(n)$ loop model and the dilute $C_2^{(1)}$ loop model. 
This list also includes the Ashkin-Teller model \cite{LR,IR}. 

As yet there is no satisfactory explanation for this unexpected 
connection between discrete holomorphicity and Yang-Baxter 
integrability. Our aim in this paper is to explicitly establish that 
the star-triangle equation, and thus Yang-Baxter integrability, is a
consequence of discrete holomorphicity. We will do this in the context
of general rhombi on a Baxter lattice and the $\ZN$ model. Our 
approach is thus algebraic. 

The utility and power of the Yang-Baxter equation is well known. 
The importance and full power of discrete holomorphicity is becoming 
increasingly recognised. Setting up a discretely holomorphic observable
has been a very useful step for the passage from discrete lattice
models to the continuous functions of conformal field theory
\cite{S,DSlectures}. Among other examples, discrete holomorphicity 
is a key ingredient in Duminil-Copin and Smirnov's \cite{DS} remarkable
and long sought after rigorous proof of the connective constant for
self-avoiding walks on the honeycomb lattice. 
The exact value had been obtained 30 years ago by Nienhuis
\cite{Nienhuis} in the $n = 0$ limit of the $O(n)$ loop model on the
honeycomb lattice. Nienhuis used various mappings between the vertex
weights of different lattice models. The key point is that for $n=0$
all contributions from closed loops vanish, leaving only self-avoiding
walks.
The crucial first step in the arguments used by Duminil-Copin and
Smirnov is the introduction of a parafermionic observable on the 
lattice which is discretely holomorphic. 
The second step involved the use of counting arguments in a domain of
the hexagonal lattice. This step built on previous rigorous 
mathematical results obtained for decomposition of excursions across
finite domains.  A more formal proof has been given by Klazar
\cite{klazar}.
The fact that such exact and rigorous results exist is no doubt related 
to the underlying integrability of the $O(n)$ loop model on the
honeycomb lattice \cite{Baxter}.

The arguments used by Duminil-Copin and Smirnov \cite{DS} were extended
to the $O(n)$ loop model on the honeycomb lattice with a boundary by 
Beaton, Bousquet-Melou, de Gier, Duminil-Copin and Guttmann \cite{BDG}
to give a mathematically rigorous proof for the critical surface 
adsorption temperature of self-avoiding walks  on the honeycomb lattice. 
This polymer adsorption transition corresponds to the $n = 0$ limit 
of the $O(n)$ model at the special surface transition. 
The exact value for the adsorption transition had been obtained earlier 
by Batchelor and Yung  \cite{BY} from the boundary Boltzmann weights
obtained by solving the boundary version of the Yang-Baxter equation.
Discrete holomorphicity at a boundary has been considered by Ikhlef
\cite{I} who was able to recover the  boundary weights of the $O(n)$
loop model on the square lattice and to obtain a new set of boundary
weights for the $\ZN$ model.

\section{From discrete holomorphicity to the star-triangle equation}

For a $\ZN$ symmetric $N$-state spin model on a graph, the weight
$w(s_a, s_b)$ of the interaction on an edge $\edge{a}{b}$ is 
unchanged by the global transformations $s_r \mapsto \omega s_r$
and $s_r \mapsto s_r^*$ of the spins. Here the spins take values
from the $N$th roots of unity, $s_r = \omega^{q_r}$, with
$q_r \in \{0, 1, \ldots, N-1\}$ and 
$\omega = \exp\wrap{{2\pi\ii/N}}$. For a review of these models
and other important models contained therein, such as the 
Ising and Potts models, see, e.g., \cite{Wu}.

Since the weight depends only
on the difference of the $q_r$ variables, we write it as $W(q_a - q_b)$,
or, by a slight abuse of notation, simply $W(a-b)$. The 
Kramers-Wannier duality of the model \cite{WuWang} lets one define
disorder variables $\disorder{r}$ dual to the spins on the dual graph. The effect of
inserting  a disorder variable at the dual site $\tilde{r}$ is to
introduce a string connecting the site to a site at infinity or at 
the boundary. Whenever the string intersects an edge of the original
lattice, it modifies the weight on that edge by, for definiteness,
lowering the spin $q_r$ to its right by one. 

\begin{figure}[h]
\centering \includegraphics{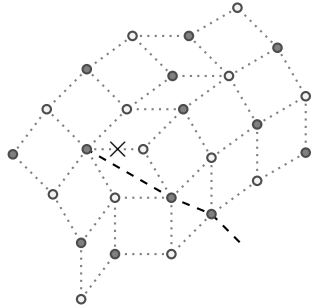}
\caption{A rhombic embedding of the covering lattice of a 
heterogeneous model showing a string attached to a disorder variable.
The empty circles denote the spin variables, while the full circles
denote the disorder variables. The location of the parafermion is 
marked with an $\times$.
}
\label{fig:fig1}
\end{figure}

\begin{figure}[h]
\centering \includegraphics{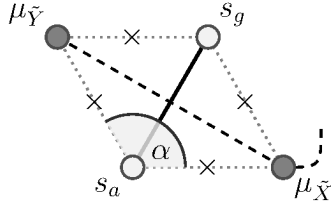}
\caption{An elementary rhombus with opening angle $\alpha$. The string
enters the rhombus via the disorder variable $\disorder{X}$. The
arbitrary axis is in the direction of the edge $\edge{a}{X}$. }
\label{fig:fig2}
\end{figure}

We consider a rhombic embedding of the covering lattice, the union
of the original lattice and its dual, onto the complex plane (figure 
\ref{fig:fig1}). 
Each elementary rhombus of this lattice contains an edge from the
original lattice and an edge from the dual lattice. The parafermion
introduced by  Rajabpour and Cardy \cite{RC}
\begin{equation*}
\RC{r}{r} = \exp\wrap{-\ii\,\sigma\,\theta_{\coveredge{r}{r}}}
\cdot s_r\cdot \disorder{r}
\end{equation*}
lives on the midpoint of the edges. Here, the angle 
$\theta_{\coveredge{r}{r}}$ is the angle between the edge traversed 
from the  spin to the disorder variable and an arbitrary but fixed 
axis. The real parameter $\sigma$ can be identified as the 
conformal spin of the observable. When the string enters a particular rhombus 
of opening angle $\alpha$ through the disorder variable $\disorder{X}$
(figure \ref{fig:fig2}), imposing the condition of discrete
holomorphicity
\begin{equation}
\sum_{\lozenge} \RC{r}{r}\,\Delta z_{\coveredge{r}{r}} = 0
\end{equation}
on the contour sum of the parafermion around the rhombus in the 
counterclockwise direction we get,
\begin{equation*}
-\ee^{\ii\,\sigma\,\pi}\,s_g\,\disorder{Y}
-\ee^{\ii\,(1-\sigma)\,\alpha}\,s_a\,\disorder{Y}
+s_a\,\disorder{X}
+\ee^{\ii\,\sigma\,\pi}\,\ee^{\ii\,(1-\sigma)\,\alpha}
\,s_g\,\disorder{X} = 0
\end{equation*}
where the variables $s_a$, $\disorder{X}$, $s_g$, $\disorder{Y}$ are at
the corners of the contour, in that order, and $\Delta z$ is the difference between the complex coordinates of the two endpoints of the side of the 
rhombus on which the $\RC{r}{r}$ variable lives, traversed in the direction of traversal of the contour.

As was explained in \cite{RC, I}, we have taken the special case $m = 1$
of the more general definition of the parafermion 
\begin{equation*}
\psi^{(m)}_{\coveredge{r}{r}}=\exp\wrap{-\ii\,\sigma_{m}\,\theta_{\coveredge{r}{r}}}
\cdot s^m_r\cdot \mu^{(m)}_{\tilde{r}}
\end{equation*}
where the string attached to the disorder variable 
$\mu^{(m)}_{\tilde{r}}$ modifies the weight of an edge it crosses by
lowering the spin to its right by $m$, and $m \in \{1, \ldots, N-1\}$.
These observables, which are also discretely holomorphic at integrable
critical points, are related to the one we consider by the global
transformation $s_r \mapsto {s^m_r}$, and thus, it is enough to
consider the $m=1$ case only without loss of generality.

With the introduction of 
$\phi(\alpha) = \ee^{\ii\,(1-\sigma)\,\alpha}$, the condition takes 
the form
\begin{equation} \label{eq:onerhombus}
\phi(-\pi)\,s_g\,\disorder{Y}
-\phi(\alpha)\,s_a\,\disorder{Y}
+s_a\,\disorder{X}
-\phi(\alpha-\pi)\,s_g\,\disorder{X} = 0
\end{equation}
Since, by our characterization of the disorder variables,
\begin{equation*}
\frac{\disorder{Y}}{\disorder{X}} 
= \frac{W_{\alpha}(a-(g-1))}{W_{\alpha}(a-g)} 
= \frac{W_{\alpha}(n_a+1)}{W_{\alpha}(n_a)}
\end{equation*}
we have
\begin{equation}\label{eq:recurrence}
\wrap{\phi(-\pi)-\phi(\alpha)\,\omega^{n_a}} W_{\alpha}(n_a+1)
+\wrap{\omega^{n_a}-\phi(\alpha-\pi)} W_{\alpha}(n_a)
 = 0
\end{equation}
which is a linear recurrence relation in the weights in terms of the
bond varible $n_a$. 

This relation specifies the weights up to a physically irrelevant
constant. The $\ZN$ symmetries, however, put constraints on the value of
the unknown $\sigma$ (see, e.g., \cite{I}). Since the weights are
symmetric and periodic, we must have 
$W_{\alpha}(n_a) = W_{\alpha}(N-n_a)$
for every $n_a$, from which the recurrence relation gives
\begin{equation*}
\wrap{1-\phi(2\alpha)}\wrap{\omega^{-1}- \phi(-2\pi)} = 0
\end{equation*}
Taking
\begin{equation}\label{eq:constraint}
\phi(2\pi) = \omega
\end{equation}
makes the parafermion holomorphic for every opening angle $\alpha$. 
Therefore,
\begin{equation}
\sigma = 1 - \frac{1}{N} + \ell
\end{equation}
for any integer $\ell$. The case $\ell = 0$ gives the 
Fateev-Zamolodchikov solutions \cite{FZ}. Since the subsequent
calculations make use of relation (\ref{eq:constraint}) and not
the actual value of $\sigma$, the proof presented here is more 
general than merely proving that the Fateev-Zamolodchikov weights 
give us an integrable model. The angle $\alpha$ is proportional to 
(in fact, in \cite{FZ}, equal to) the spectral parameter and plays 
an analogous r\^ole.

It should be noted that the weights are real because the ratio of 
the weights given by the recurrence relation (\ref{eq:recurrence})
is real.

\subsection{Self-duality}

\begin{figure}[h]
\centering \includegraphics{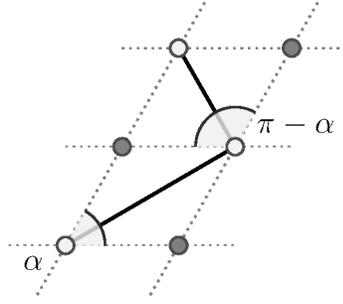}
\caption{Rhombic embedding of an anisotropic square lattice. The
edge with an opening angle $\alpha$ is horizontal, while that
with an opening angle $\pi-\alpha$ is vertical in the original
lattice.}
\label{fig:fig3}
\end{figure}

We now show that the weights are self-dual. For this, we temporarily
consider the embedding of an anisotropic square lattice (figure
\ref{fig:fig3}). On the one hand, if the horizontal interaction edges 
are mapped onto rhombi of opening angle $\alpha$, the vertical edges 
are then mapped onto rhombi with opening angle $\pi-\alpha$. We thus
have $\overline{W}_{\alpha}(n_a) = W_{\pi-\alpha}(n_a)$. On the other
hand,  taking the discrete Fourier transform of the recurrence 
relation (\ref{eq:recurrence}) we have
\begin{equation*}
\wrap{\phi(-\pi)-\phi(-\pi-\alpha)\,\omega^{k_a+1}}
\widetilde{W}_{\alpha}(k_a+1)
+ \wrap{\omega^{k_a} - \phi(-\alpha)}
\widetilde{W}_{\alpha}(k_a)
= 0
\end{equation*}
Using (\ref{eq:constraint}) to replace the extra $\omega$ in the first
term, we see that the dual weights satisfy the same equation with
$\alpha$ replaced by $\pi-\alpha$. Since our unitary Fourier transform
preserves the norm 
\begin{equation}
\sum_{n_a\in \ZN}[W_{\alpha}(n_a)]^2 = 1
\end{equation} this proves
\begin{equation}
\overline{W}_{\alpha}(n_a) = W_{\pi-\alpha}(n_a) 
= \widetilde{W}_{\alpha}(n_a)
\end{equation}
and captures the crossing symmetry of the problem.

\subsection{Inversion relations}
\begin{figure}[h]
\centering \includegraphics{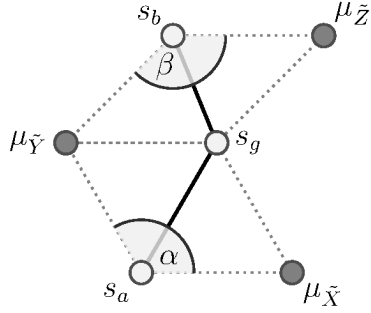}
\caption{Two adjacent rhombi. }
\label{fig:fig4}
\end{figure}

Now we consider another rhombus with angle $\beta$ which sits on top of
the rhombus considered so far (figure \ref{fig:fig4}). The new rhombus
comes with spin $s_b$ and disorder variable $\disorder{Z}$ and shares
the edge $\coveredge{g}{Y}$ with the previous one. We multiply its
holomorphicity equation analogous to (\ref{eq:onerhombus}) by
$\phi(-\beta)$ to orient it correctly and then add the two equations 
to get
\begin{eqnarray}\label{eq:tworhombi}
-\phi(\alpha)\,s_a\,\disorder{Y}
+s_a\,\disorder{X}
-\phi(\alpha-\pi)\,s_g\,\disorder{X} \nonumber\\
\quad\quad+ \,\phi(-\pi-\beta)\,s_g\,\disorder{Z}
- s_b\,\disorder{Z}
+\phi(-\beta)\,s_b\,\disorder{Y}
= 0
\end{eqnarray}
The contributions from the common edge cancel as it is traversed
in opposite directions for the two sums. Using
\begin{equation*}
\frac{\disorder{Z}}{\disorder{Y}} 
= \frac{W_{\beta}(b - (g-1))}{W_{\beta}(b-g)} 
= \frac{W_{\beta}(n_b + 1)}{W_{\beta}(n_b)} 
\end{equation*}
the equation (\ref{eq:tworhombi}) gives a quadratic relation 
\begin{eqnarray}
\wrap{ \omega^{a}- \phi(\alpha-\pi)\,\omega^{g}}W_{\alpha}(n_a)
\,W_{\beta}(n_b) 
\nonumber\\ \quad\quad
+\, \wrap{\phi(-\beta)\,\omega^{b}-\phi(\alpha)\,\omega^{a}}
W_{\alpha}(n_a + 1)\,W_{\beta}(n_b) 
\nonumber\\ \quad\quad
+\,\wrap{\phi(-\pi-\beta)\,\omega^{g}-\omega^{b}}W_{\alpha}(n_a + 1)
\,W_{\beta}(n_b+1)
=0
\end{eqnarray}
in the weights. 

This relation contains the inversion relations 
(see, e.g., \cite{Z}) for the model. 
Put $a = b$, so that $n_a = n_b = n$, and $\beta = -\alpha$, then 
the equation reduces to
\begin{eqnarray*}
W_{\alpha}(n)\,W_{-\alpha}(n) = W_{\alpha}(n + 1)\,W_{-\alpha}(n+1)
\end{eqnarray*}
thus, the product is independent of $n$, 
\begin{equation} \label{eq:ir1}
W_{\alpha}(n)\,W_{-\alpha}(n) = g(\alpha)\,g(-\alpha)
\end{equation}
The other relevant inversion relation
\begin{equation}  \label{eq:ir2}
\sum_{g\in\ZN}W_{\pi+\theta}(a-g)\,W_{\pi-\theta}(b-g)
=\rho(\theta)\,\delta_{a,b}
\end{equation}
follows from this relation and self-duality by expanding the left hand
side in Fourier coefficients and gives
\begin{equation*}
\rho(\theta) = N\,g(\theta)\,g(-\theta)
\end{equation*}
for our normalization.

\subsection{Star-triangle relation}

\begin{figure}[h]
\centering \includegraphics{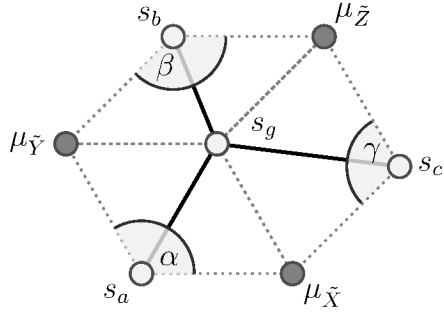}
\caption{Three adjacent rhombi making the star. }
\label{fig:fig5}
\end{figure}

We proceed to add the third rhombus with angle 
$\gamma = 2\pi - (\alpha+\beta)$ to make the star (figure 
\ref{fig:fig5}). Two of the edges, $\coveredge{g}{Z}$ and 
$\coveredge{g}{X}$, are shared and a new spin variable $s_c$ is
introduced. Multiplying the holomorphicity equation analogous to
(\ref{eq:onerhombus}) for this rhombus by  $\phi(2\pi -\alpha)$ to
orient it correctly, we add it to the contour sum (\ref{eq:tworhombi})
to yield 
\begin{eqnarray}
\label{eq:threerhombi}
\phi(\alpha-3\pi)\,s_g\,\bar{\mu}_{\tilde{X}}
-\phi(\alpha-\pi)\,s_g\,\disorder{X}
\nonumber\\ \quad\quad
+\, s_a\,\disorder{X}-\phi(\alpha)\,s_a\,\disorder{Y} 
\nonumber\\ \quad\quad
+\,\phi(-\beta)\,s_b\,\disorder{Y}-s_b\,\disorder{Z}
\nonumber\\ \quad\quad
+\,\phi(\alpha - 2\pi)\,s_c\,\disorder{Z}
-\phi(-\beta)\,s_c\,\bar{\mu}_{\tilde{X}} = 0
\end{eqnarray}
Some features of this equation warrant attention. The disorder variable
$\disorder{X}$ is not the same as $\bar{\mu}_{\tilde{X}}$, as might
be na\"ively expected. A possible interpretation of this is that
the string has gone around the spin $s_g$ once, and thus the
parafermion has acquired a phase factor $\phi(2\pi)$, corresponding
to the lowering of the value of $g$, that is, $q_g$, by one in the 
configuration. Hence, the contributions from the edge 
$\coveredge{g}{X}$ do not cancel.

Making use of
\begin{equation*}
\frac{\bar{\mu}_{\tilde{X}}}{\disorder{Y}} 
= \frac{W_{\gamma}(c-(g-1))}{W_{\gamma}(c-g)} 
= \frac{W_{\gamma}(n_c+1)}{W_{\gamma}(n_c)}
\end{equation*}
we have a cubic equation in weights which we omit for clarity. 
However, summing over $g$, we arrive at
\begin{eqnarray}\label{eq:LHSrecursion}
\wrap{\omega^{a}-\phi(-\beta)\,\omega^{c}}
\,\mathfrak{L}_{\alpha,\beta}(a,b,c)
\nonumber\\ \quad\quad
+\,\wrap{\phi(-\beta)\,\omega^{b}-\phi(\alpha)\,\omega^{a}}
\,\mathfrak{L}_{\alpha,\beta}(a+1,b,c)
\nonumber\\ \quad\quad
+\,\wrap{\phi(\alpha-2\pi)\,\omega^{c}- \omega^{b}}
\,\mathfrak{L}_{\alpha,\beta}(a+1,b+1,c) = 0
\end{eqnarray}
Here 
\begin{equation}
\mathfrak{L}_{\alpha,\beta}(a,b,c) 
=\sum_{g\in\ZN}W_{\alpha}(a-g)\,W_{\beta}(b-g)\,W_{\gamma}(c-g)
\end{equation}
is the left hand side of the star-triangle relation. The terms 
involving $s_g$ cancel because of the constraint (\ref{eq:constraint})
on $\sigma$, and therefore, the expectation value of the parafermion 
is still independent of the path, as is that for the disorder variable
$\disorder{X}$, and consequently the contour sum around the star still
vanishes.

\begin{figure}[h]
\centering \includegraphics{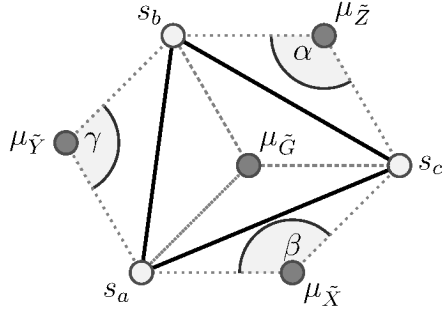}
\caption{Three adjacent rhombi making the triangle. }
\label{fig:fig6}
\end{figure}

Remarkably, considering the contour sum around the triangle (figure
\ref{fig:fig6}), we get
\begin{eqnarray}\label{eq:RHSrecursion}
\wrap{\omega^{a}-\phi(-\beta)\,\omega^{c}}
\,\mathfrak{R}_{\alpha,\beta}(a,b,c)
\nonumber\\ \quad\quad
+\,\wrap{\phi(-\beta)\,\omega^{b}-\phi(\alpha)\,\omega^{a}}
\,\mathfrak{R}_{\alpha,\beta}(a+1,b,c)
\nonumber\\ \quad\quad
+\,\wrap{\phi(\alpha-2\pi)\,\omega^{c}- \omega^{b}}
\,\mathfrak{R}_{\alpha,\beta}(a+1,b+1,c) = 0
\end{eqnarray}
where 
\begin{equation}
\mathfrak{R}_{\alpha,\beta}(a,b,c) 
= W_{\pi-\gamma}(a-b)\;W_{\pi-\alpha}(b-c)\;W_{\pi-\beta}(c-a)
\end{equation}
is proportional to the right hand side of the star-triangle relation.
The two sides of the equation therefore obey the same recurrence
relation.

To show that the star-triangle relation follows from these relations, 
we have to prove that the two solutions are linearly dependent. 
To this end, we consider a general solution of the
recurrence relation $f_{\alpha,\beta}(a, b, c)$. 
Taking the complex conjugate of the relation and noting both 
$f_{\alpha,\beta}(a, b, c)$ and $\sigma$ are real, we multiply the
resulting equation by $\phi(\alpha-\beta)\,\omega^{a+b}$. The result,
\begin{eqnarray}
\wrap{\phi(\alpha-\beta)\,\omega^{b}-\phi(\alpha)\,\omega^{a+b-c}}
\,f_{\alpha,\beta}(a,b,c)
\nonumber\\ \quad\quad
+\,\wrap{\phi(\alpha)\,\omega^{a}-\phi(-\beta)\,\omega^{b}}
\,f_{\alpha,\beta}(a+1,b,c)
\nonumber\\ \quad\quad
+\,\wrap{\phi(2\pi-\beta)\,\omega^{a+b-c}
  - \phi(\alpha-\beta)\,\omega^{a}}
\,f_{\alpha,\beta}(a+1,b+1,c) = 0
\end{eqnarray}
when added to the recurrence relation, cancels the middle term 
and gives the ratio
\begin{equation}
\frac{f_{\alpha,\beta}(a+1,b+1,c)}{f_{\alpha,\beta}(a,b,c)} 
= \frac{f_{\alpha,\beta}(a,b,c-1)}{f_{\alpha,\beta}(a,b,c)}
\end{equation}
independent of the function $f_{\alpha,\beta}$. The shift in the
arguments is justified by the $\ZN$ symmetries. Since this ratio is 
the same for both $\mathfrak{R}_{\alpha,\beta}$, 
and $\mathfrak{L}_{\alpha,\beta}$,
\begin{equation}
\frac{\mathfrak{L}_{\alpha,\beta}(a,b,c)}
{\mathfrak{R}_{\alpha,\beta}(a,b,c)} 
= \frac{\mathfrak{L}_{\alpha,\beta}(a,b,c-1)}
{\mathfrak{R}_{\alpha,\beta}(a,b,c-1)}
\end{equation}
The ratio of the two sides is thus independent of the spin $c$. Similar
elimination of the other two terms in the recurrence relation shows that
the ratio is independent of $a$ and $b$ as well, as is demanded by
symmetry.\footnote{Jacques Perk informed us that this style of argument 
has been used to prove that the Boltzmann weights 
of the more general chiral Potts model satisfy the star-triangle relation \cite{Perk}.}
We therefore have the star-triangle relation 
\begin{equation} \label{eq:str}
\mathfrak{L}_{\alpha,\beta}(a,b,c) 
= R_{\alpha,\beta}\, \mathfrak{R}_{\alpha,\beta}(a,b,c)
\end{equation}
where $R_{\alpha,\beta}$ depends on the spectral variables only.

As is well-known, the inversion relations (\ref{eq:ir1})-(\ref{eq:ir2}) 
and the star-triangle relation (\ref{eq:str}) together are sufficient to 
establish the existence of commuting transfer matrices parametrized by, 
in our case, the angle of the embedded rhombi, or in other words, 
Yang-Baxter integrability. By the embedding onto the complex plane,
the schematic diagram of the Yang-Baxter equations acquires a geometric
meaning of rearrangements of the elementary rhombi. Our analysis
shows that the contour sum around a domain picks up only factors 
independent of the configuration under such rearrangements.

\subsection{Homogeneous lattices}
\begin{figure}[h]
\centering \includegraphics{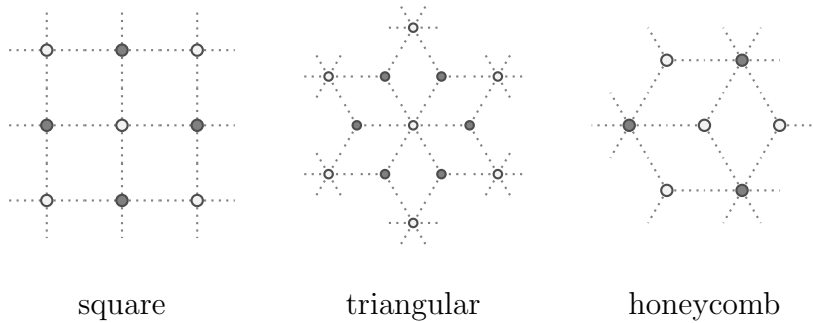}
\caption{Rhombic embedding of the three homogeneous lattices.}
\label{fig:fig7}
\end{figure}

It is interesting to note that the construction presented above
is local in that it does not depend on many of the properties,
e.g., the coordination number, of the underlying graph. To see
the consequences of this lattice independence, we consider the 
three archetypical homogeneous lattices, the square, the triangular 
and the honeycomb (figure \ref{fig:fig7}). The embedding of these
lattices onto the complex plane tiles the plane with $\alpha$ being
$\pi/2$, $\pi/3$ and $2\pi/3$ respectively. Evaluating the recurrence
relation (\ref{eq:recurrence}) at these points gives an instant
derivation of the known critical weights. For example, for the 
Ising model, we have for $W(1)/W(0)$, the known values $\sqrt{2} - 1$, 
$1/\sqrt{3}$, and $2 - \sqrt{3}$ for the three lattices respectively
\cite{BaxterBook}.

\section{Concluding remarks}

In this paper, we considered the implication of the condition of
discrete holomorphicity on two and three adjacent rhombi in the context
of the lattice $\ZN$ model. For two rhombi this led to the quadratic 
equation (\ref{eq:tworhombi}) in the Boltzmann weights. This equation
was shown to imply the known inversion relations (\ref{eq:ir1}) and
(\ref{eq:ir2}) for this model. Note that Cardy \cite{Cardy-H}  
has effectively established the inversion relations in the $u-v \to 0$
limit of the Yang-Baxter equation. 
For three rhombi we obtained the cubic equation (\ref{eq:threerhombi})
in the Boltzmann weights. In establishing this equation the lattice
parafermion picks up a crucial phase factor, with the expectation value
of the parafermion still independent of the path. The importance of such
a phase factor has been highlighted in the topological context of the
loop models by Fendley \cite{F}. Here we have shown that the 
star-triangle relation (\ref{eq:str}) follows from the three-rhombus
equation (\ref{eq:threerhombi}). 
In the discrete holomorphic approach the two-rhombus equation
(\ref{eq:tworhombi}) and the three-rhombus equation
(\ref{eq:threerhombi})  can thus be considered as analogues of the 
two- and three-body conditions for integrability. 
However, the simplicity of the discrete holomorphic approach is that
ultimately these conditions are equivalent to the one-rhombus equation
(\ref{eq:onerhombus}). 
Indeed, one can push this argument further by building up a transfer
matrix from the rhombi and showing that commuting transfer matrices can
be established as a consequence of discrete holomorphicity, bypassing
the use of the Yang-Baxter equation. 
Our results lend further impetus for using discrete holomorphicity as 
a tool for investigating lattice models at critical points.

\ack It is a pleasure to dedicate this paper to Fred Wu on the occasion
of his 80th birthday. We thank Vladimir Bazhanov for a number of helpful discussions and 
Jacques Perk for a helpful remark. 
This work has been partially supported by the Australian Research Council.

\end{document}